\documentclass[epj,nopacs,final]{svjour}
\usepackage{graphicx}
\usepackage{amsmath}
\usepackage{upgreek}
\usepackage{amssymb}
\usepackage{latexsym}
\usepackage{epstopdf}
\usepackage{flushend}

\begin{document}

\title{Snap-off production of monodisperse droplets}
\author{Solomon Barkley\inst{1} 
\and Eric R. Weeks \inst{2} 
\and Kari Dalnoki-Veress \inst{1,3} \thanks{email: dalnoki@mcmaster.ca}
}
\institute{
Department of Physics \& Astronomy and the Brockhouse Institute for Materials Research,
McMaster University, Hamilton, ON, Canada 
\and Department of Physics, Emory University, Atlanta, GA 30322 USA 
\and PCT Lab, UMR CNRS 7083 Gulliver, ESPCI ParisTech, PSL Research University, Paris, France}


\date{\today}
\abstract{
We introduce a novel technique to produce monodisperse droplets through the snap-off mechanism. The methodology is simple, versatile, and  requires no specialized or expensive components. The droplets produced  have polydispersity $<1 \%$ and can be as small as $2.5~\mu$m radius.  A convenient feature is that the droplet size is constant over a 100-fold change in flow rate, while at  higher flows the droplet size can be continuously adjusted.
}

\authorrunning{Barkley et al.}

\maketitle

Microfluidics applications often require emulsions with a wide range of characteristics, prompting the development of several distinct techniques for producing droplets\cite{christopher07,seeman12}. One important parameter is the degree of polydispersity among droplet sizes, where smaller values are preferable for many applications. Droplets with an extremely low polydispersity are particularly desirable for basic science investigations of emulsions~\cite{baret09,WeeksForceChains}, vessels for tiny experiments \cite{stan09,selimovic10,tanaka15}, as well as calibration in both academic and industrial settings~\cite{UlmkeCalibration}. Here we present a method we have recently developed using glass capillaries and a surface tension driven `snap-off' instability to produce droplets. This method is remarkable for its simplicity, ease of implementation, and the high monodispersity of droplets produced. An additional convenience is that there are two distinct regimes of droplet production: 1) the size of droplets is insensitive at low flow rates; while, 2) at high flow rates the droplet size is tunable.

\begin{figure}[t]
\begin{center}
\includegraphics[width=65mm]{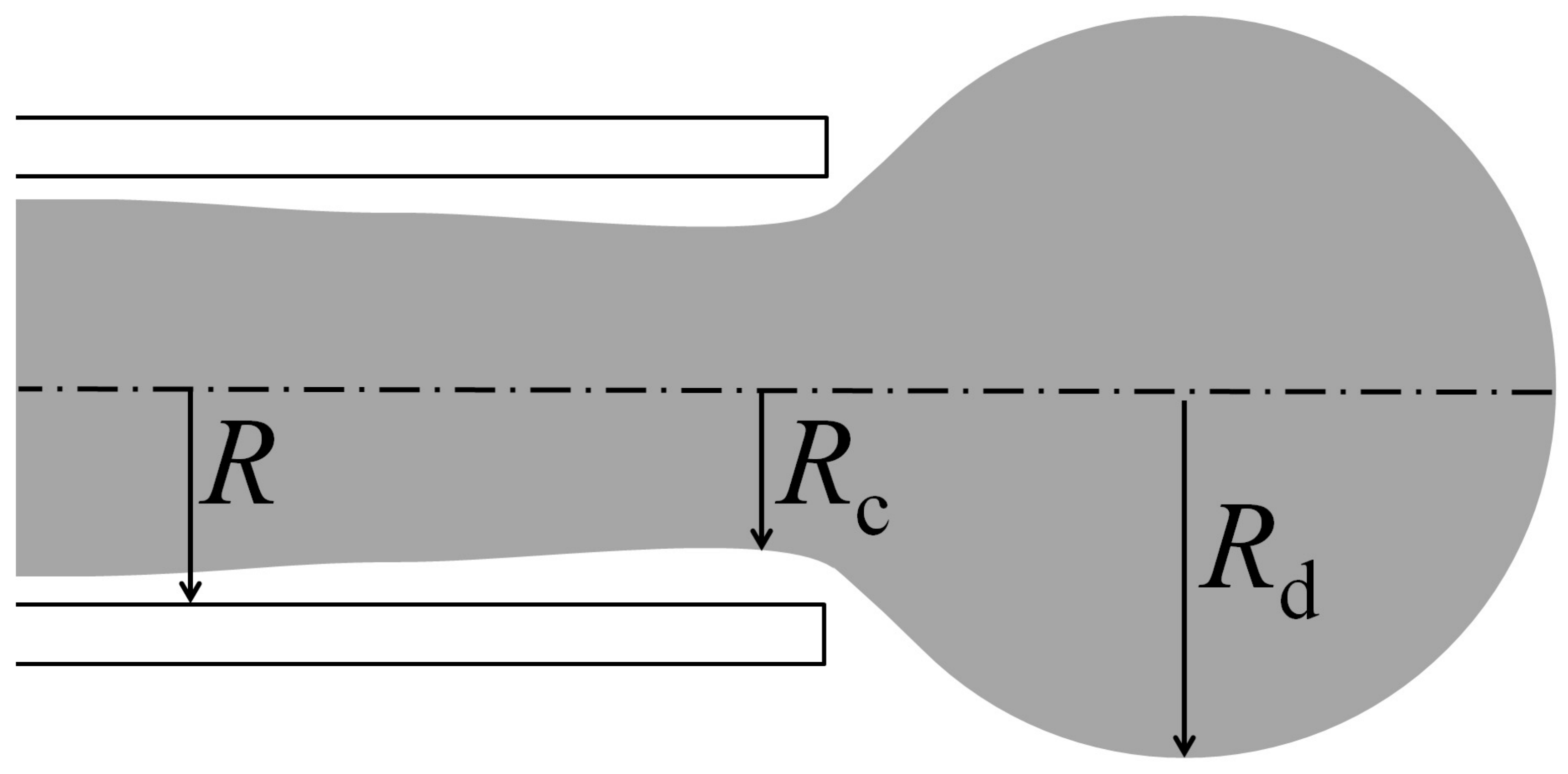}
\caption{Schematic of the experimental setup before a droplet snaps off. A column of dispersed phase (grey) with minimum radius $R_{\mathrm{c}}$ is ejected from a glass capillary with inside radius $R$ into the dispersed phase (white). The continuous phase forms a thin wetting layer along the inside of the tube. The dispersed phase forms a growing droplet of radius $R_{\mathrm{d}}$ at the end of the tube.}
\label{fig:Schematic}
\end{center}
\end{figure}

The snap-off instability of droplets in cylindrically symmetric capillaries was first described in 1970~\cite{Roof1970}, and has since been investigated further in the context of understanding the physics behind  snap-off~\cite{LenormandSnapOff,PenaSnapOff}.  We took advantage of this effect to develop a versatile system for production of monodisperse droplets that is easy to assemble and operate. One important consideration is that this setup requires no flow of the continuous phase, since the pinch-off is driven by surface tension forces rather than viscous forces. Although the snap-off process has been used previously to produce droplets in flattened microfluidic geometries~\cite{DanglaDroplets,vanDijkeDroplets,malloggi10}, our simple cylindrical configuration is able to produce droplets that are more monodisperse.  

In order to prepare monodisperse droplets we have utilised a method that is schematically depicted in Fig.~\ref{fig:Schematic}. A thin glass capillary tube of radius $R$ is filled with the dispersed phase which is ejected with some flow rate into the continuous phase. A growing droplet forms at the end of the tube with radius $R_{\mathrm{d}}$. At low flow rates we can treat the system as being quasi-static. This dictates that the dispersed phase immediately inside the tube has the same pressure as that in the droplet~\cite{Roof1970,vanDijkeDroplets}. However, due to Laplace pressure, the pressure of the continuous phase surrounding the dispersed phase is not uniform and varies with axial position. The Laplace pressure decrease between the dispersed phase and continuous phase is $\Delta P_{\mathrm{d}}=2\gamma/R_{\mathrm{d}}$ for the droplet, with interfacial tension $\gamma$~\cite{PGG}. If the continuous phase wets the capillary, then a thin wetting layer coats the inside surface of the tube. We define the collar to be the point at which the radius of the dispersed phase within the tube is a minimum,  $R_{\mathrm{c}} \le R$. Since the radius at the collar is much smaller than the orthogonal radius, we can neglect the orthogonal contribution to the Laplace pressure at the collar and write the Laplace pressure decrease at the collar as $\Delta P_{\mathrm{c}}=\gamma/R_{\mathrm{c}}$.  Using the fact that the pressure in the dispersed phase is the same at the collar and in the droplet, the pressure difference between the continuous phase at the collar and that in the bulk is given by the difference in these two Laplace pressures:
\begin{equation}
\Delta P=\Delta P_{\mathrm{d}}-\Delta P_{\mathrm{c}}=\gamma\left(\frac{2}{R_{\mathrm{d}}}-\frac{1}{R_{\mathrm{c}}}\right).
\label{eq:QuasiStatic}
\end{equation}
It follows that if $\Delta P<0$, the continuous phase will be drawn into the tube around the dispersed phase and the wetting layer thickens ($R_{\mathrm{c}}$ decreases and thus $\Delta P$ becomes more negative). Conversely the continuous phase will be ejected from the tube if $\Delta P>0$.

\begin{figure}
\begin{center}
\includegraphics[width=75mm]{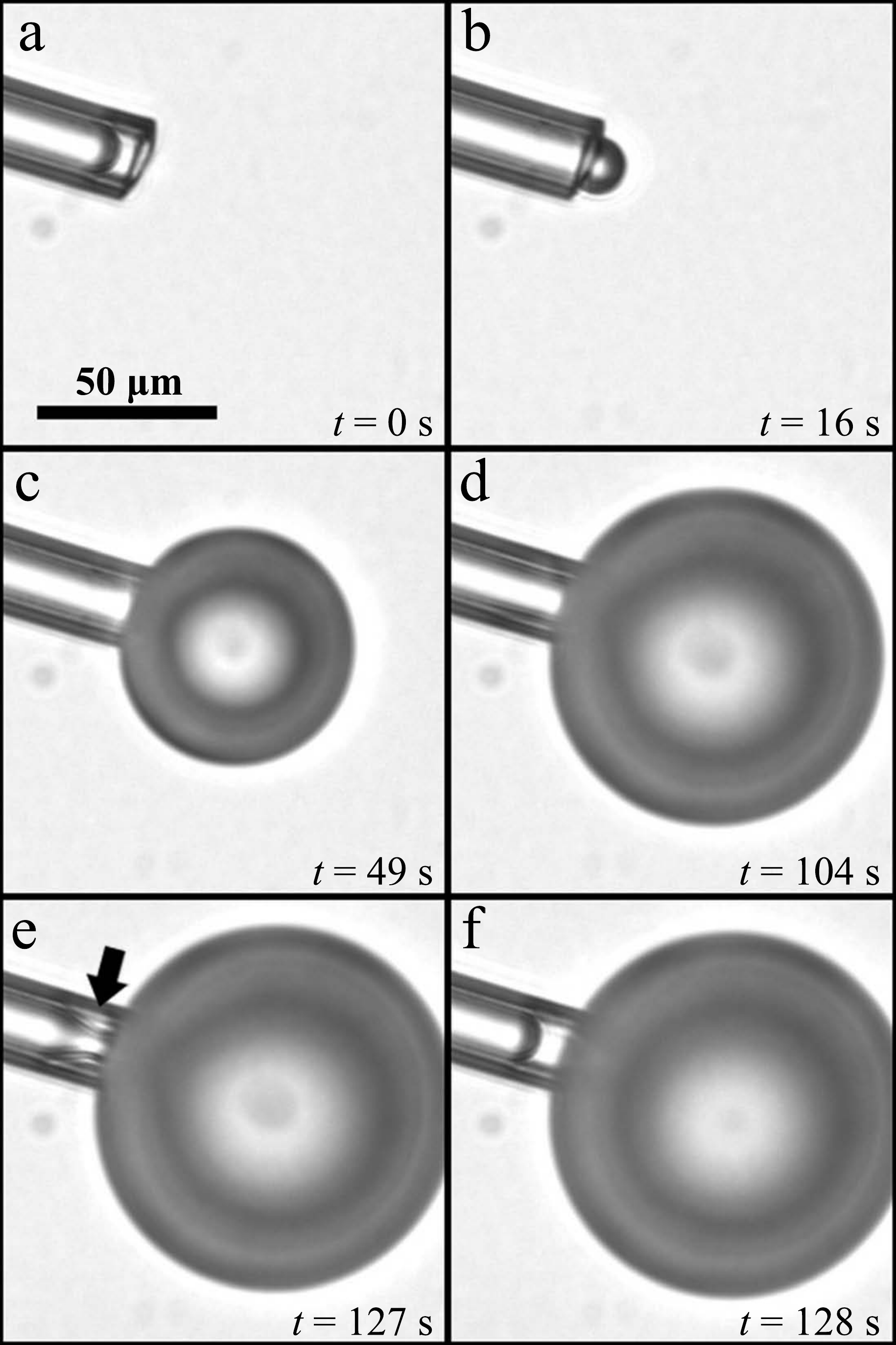}
\caption{Time series of a droplet growing until snap-off. a-c) Mineral oil is ejected from the tip of a glass pipette into water, forming a growing droplet. d) Water is spontaneously drawn into the end of the pipette once the droplet has reached a critical size. e) The water forms a collar around the column of oil flowing outwards. f) The collar constricts and snaps off the droplet. We note that the timescales over which the images were taken in this sequence is much longer than typical. The slow rate was required in order to image the rapidly forming collar structure.}
\label{fig:DropletGrowth}
\end{center}
\end{figure}

To illustrate the formation of the droplets we show a sequence of optical microscopy images of the snap-off process in Fig.~\ref{fig:DropletGrowth}. Initially, the dispersed phase is contained entirely within the tube (Fig.~\ref{fig:DropletGrowth}a) and the wetting layer of the continuous phase is very thin along the entire length of the tube ($R_{\mathrm{c}}\sim R$). Upon ejection, the dispersed phase forms a very small droplet (Fig.~\ref{fig:DropletGrowth}b). At this point the Laplace pressure in the droplet, $\Delta P_{\mathrm{d}}$, is high relative to that within the tube due to the high curvature of the small droplet. Thus there exists a positive pressure difference $\Delta P$ of the continuous phase between the inside and outside of the tube (see Eq.~\ref{eq:QuasiStatic}). As the droplet grows (Fig.~\ref{fig:DropletGrowth}c), both $\Delta P_{\mathrm{d}}$ and $\Delta P$ decrease as a result of increasing $R_{\mathrm{d}}$, until $\Delta P_{\mathrm{d}} < \Delta P_{\mathrm{c}}$ , or equivalently $\Delta P < 0$. Once $\Delta P < 0$ the continuous phase begins to invade the tube (visible as darkening in Fig.~\ref{fig:DropletGrowth}d): the configuration is unstable and there is a spontaneous reverse flow of the continuous phase. The continuous phase forms a collar around the dispersed phase which is marked by an arrow in Fig.~\ref{fig:DropletGrowth}e. During this phase $R_{\mathrm{c}}$  and $\Delta P$ decrease rapidly causing the collar to collapse -- the `snap-off' of the droplet is complete and the process repeats (Fig.~\ref{fig:DropletGrowth}f).

In the experiments presented the tube was a thin glass pipette.  Pipettes were prepared using a pipette puller (Narishige, Japan) from glass capillaries (World Precision Instruments, USA) with initial outside diameter 1~mm and inside diameter $580~\mu$m. The pipettes used had an outside diameter ranging from $<3-1000~\mu$m. We note that we have also prepared a similar device simply by heating a capillary tube over an ethanol flame and stretching it manually -- a pipette puller is merely a convenience. Unless otherwise specified the continuous phase is water and the dispersed phase is mineral oil. Droplets were stabilized against coalescence by adding surfactant ($1\%$ sodium dodecyl sulphate) to the continuous water phase. Mineral oil was forced out of the pipette by applying pressure to a syringe connected to the pipette, or by adjusting the height of an oil reservoir. As discussed, to facilitate the snap-off process there has to be a thin layer of the continuous phase surrounding the dispersed phase inside the pipette. Wetting of the inside surface of the pipette by the continuous phase was achieved by briefly reversing flow in the tube before droplet production. 

For certain applications, it may be desirable to produce water droplets in oil. This was achieved by coating the pipette with polystyrene which is wet by mineral oil as opposed to water. A solution was prepared by dissolving polystyrene of molecular weight $M_\mathrm{w}=8.8$~kg/mol in toluene at a concentration of 20\% (Polymer Source, Canada). A pipette was then coated by dipping its tip into the polystyrene solution and expelling air through the pipette as the solvent evaporated. Indeed, the crucial aspect is the wetability of the inside surface of the pipette by the continuous phase -- given that constraint the snap-off phenomenon can be made to occur for a wide range of liquids~\cite{PenaSnapOff,vanDijkeDroplets}.

\begin{figure}
\begin{center}	
\includegraphics[width=85mm]{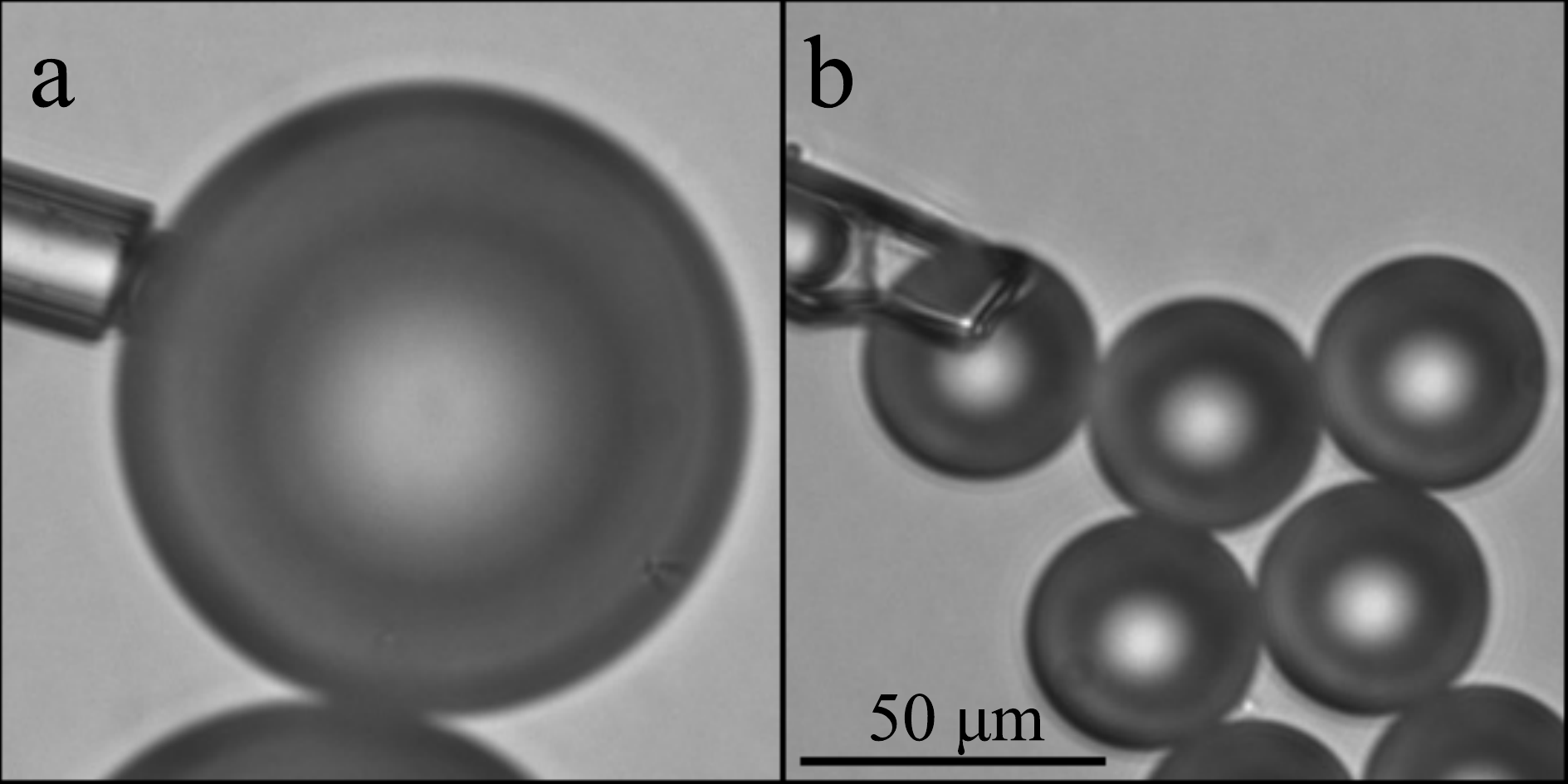}
\caption{Effect of pipette tip shape on snap-off droplet production. a)  A  tip that has a flat opening produces large droplets. b)~After the tip of the pipette shown in (a) is broken irregularly, it produces droplets that are smaller and more monodisperse.}
\label{fig:TipShape}
\end{center}
\end{figure}

Snap-off droplet size is influenced by both the diameter of the pipette and the shape of the pipette tip. By changing the radius $R$ of the pipette used to produce the droplets, it is possible to vary the radius $R_{\mathrm{d}}$ of droplets from $2.5-750~\mu$m. For snap-off to occur, the continuous phase must enter the tip of the pipette as the droplet is produced. If the pipette has a completely flat tip, as in  Fig.~\ref{fig:DropletGrowth}, the circular opening is occluded by the growing spherical droplet. As a result, the continuous phase is unable to easily flow into the pipette and snap-off is delayed.  To illustrate the effect of tip shape, a pipette that initially had a flat tip geometry is shown in Fig.~\ref{fig:TipShape}a. That tip was then broken off with a pair of tweezers to produce the irregular tip shown in Fig.~\ref{fig:TipShape}b, and the same pipette produces droplets that are both smaller and more monodisperse when it has an irregular tip shape. The irregularity of the tip facilitates the reverse flow of the continuous phase into the pipette. Eq.~\ref{eq:QuasiStatic} predicts snap-off in the quasi static regime when $\Delta P < 0$, or alternatively, when $R_{\mathrm{d}} > 2R$. This condition applies only for irregular tips where the flow of the continuous phase into the pipette is unimpeded. The importance of an irregular opening for snap-off has been demonstrated previously in a different geometry~\cite{Roof1970}.

Despite the variation in droplet size between pipettes, there is very little variation in the size of droplets produced from the same pipette at a given flow rate. A standard measure of monodispersity is the coefficient of variation (CV) in droplet radius, defined as the standard deviation in droplet radii normalized by the mean droplet radius. Pipettes with irregular tips, which are both easy to prepare and most ideal for droplet production, like that shown in Fig.~\ref{fig:TipShape}b, produce droplets with radius CV of  $0.5\%$. These droplets are more monodisperse than those produced through other recently developed methods, which all report polydispersity $>1\%$~\cite{tanaka15,DanglaDroplets,vanDijkeDroplets,malloggi10,BauerMonodisperse,DuanMonodisperse,MulliganMonodisperse,priest2006generation,chokkalingam2010optimized}. It should be noted that our measurement of droplet radius CV is limited by the precision with which we are able to measure individual droplet radii ($\sim60$~nm), and so the quoted value represents an upper bound on the actual radius CV.

\begin{figure}
\begin{center}
\includegraphics[width=85mm]{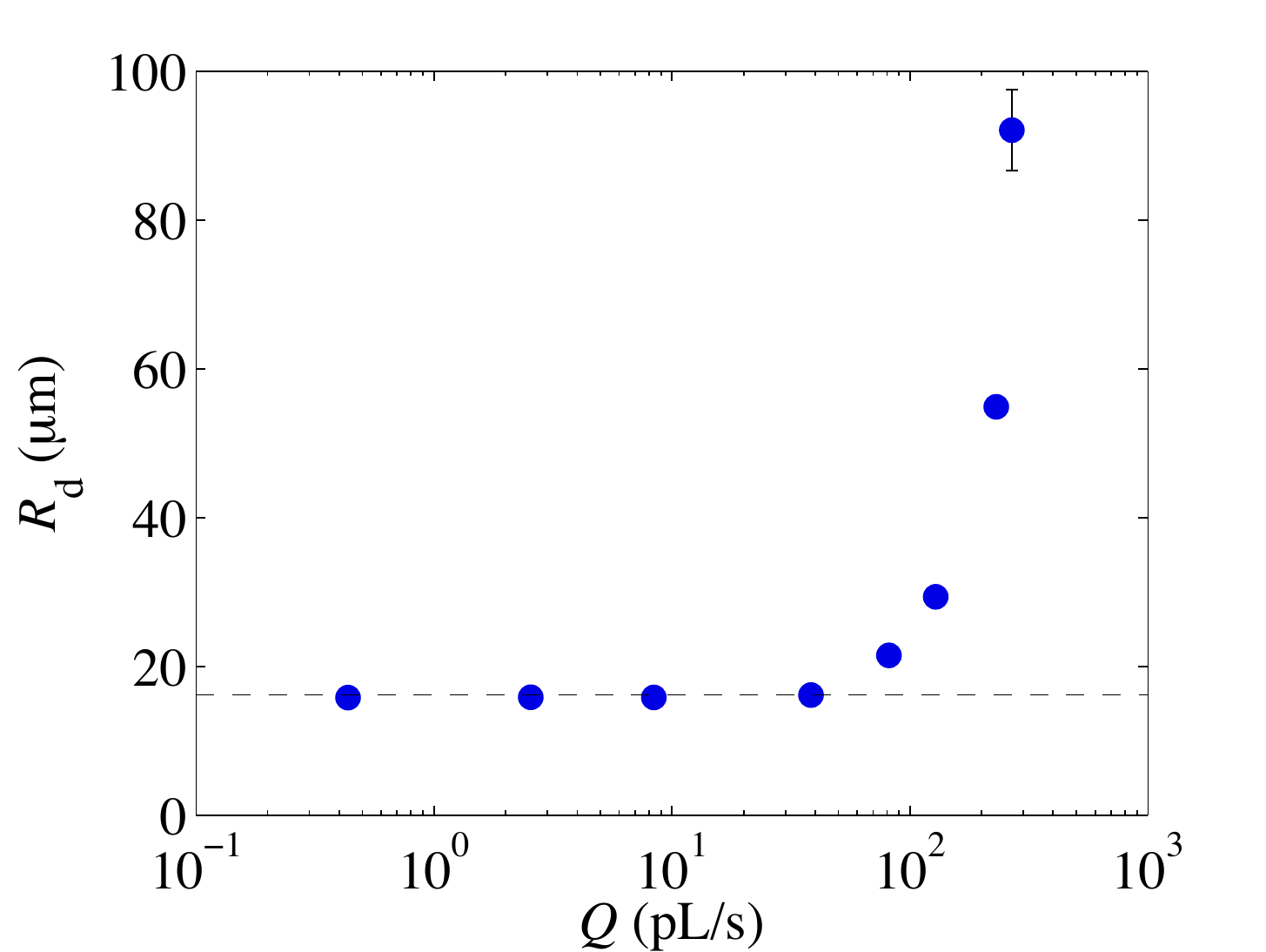}
\caption{Effect of volumetric flow rate $Q$ on droplet radius $R_{\mathrm{d}}$ for snap-off droplets produced from the same pipette. The horizontal dashed line is the quasi-static prediction from pipette diameter, $R_{\mathrm{d}}=2R$. Other than the error bar shown for the highest flow rate (mean~$\pm$ standard deviation), the error in the droplet size was much smaller than the data points due to the droplets' monodispersity.}
\label{fig:rvsQ}
\end{center}
\end{figure}

The quasi-static approximation is valid only in the limit of negligible flow of the dispersed phase from the end of the pipette (i.e. when the pressure gradient driving the flow is negligible compared to those in Eq.~\ref{eq:QuasiStatic}). At higher flow rates, snap-off droplets are found to be larger than would be predicted by Eq.~\ref{eq:QuasiStatic}. The size of droplets produced by snap-off from a single pipette $R_{\mathrm{d}}$ is shown as a function of volumetric flow rate $Q$ in Fig.~\ref{fig:rvsQ}. For $Q<50$~pL/s, droplet size is constant, and is in excellent agreement with $R_{\mathrm{d}}=2R$, as predicted by Eq.~\ref{eq:QuasiStatic} and shown in Fig.~\ref{fig:rvsQ} as the black horizontal dashed line. The quasi-static approximation is valid over a 100-fold increase in flow rate, with no measurable change in $R_{\mathrm{d}}$ up to $50$~pL/s. The insensitivity of droplet size to flow rate in this regime provides a simple approach to produce monodisperse droplets even with poor control over pressure or flow rate of the dispersed phase (i.e applying pressure to a syringe by hand).  Additionally, we find that the droplet production rate can be adjusted continuously through control of flow rates in the quasi-static regime, to a maximum of $\sim2$ droplets per second for the type of data shown in~Fig.~\ref{fig:rvsQ}.  The production rate represents one of the disadvantages of the snap-off process described here -- oftentimes higher rates of production are desirable. 

As can be seen in Fig.~\ref{fig:rvsQ}, with large flow rates the  droplet size increases monotonically. The high flux of the dispersed phase out of the pipette prevents the reverse flow of the continuous phase, and snap-off is delayed -- the quasi-static assumption is no longer valid. An added convenience of the snap-off methodology is that in this regime, droplet size is tunable by adjusting the flow rate without a need to change the radius of a pipette. In Fig.~\ref{fig:rvsQ} the size of the droplets is increased by a factor of $\sim5$ over a $\sim 10$-fold increase in flow rate with no adverse effects to droplet monodispersity. (We note that at the highest flow rate the error bar does reflect an adverse scatter in the droplet radii, and attribute this to a lack of fine control in achieving a constant flow rate at that value). The five-fold increase in droplet size is not a fundamental limitation, rather if the flow rate is increased even further, snap-off ceases entirely, and the droplet grows indefinitely~\cite{PenaSnapOff}.

In conclusion, snap-off droplet generation in pre-wet cylindrical glass capillaries is advantageous for its simplicity, versatility, and the monodispersity of produced droplets. All of the system components are readily available and no specific technical expertise is required. Droplet size can be altered by replacing the pipette or by increasing the volumetric flow rate of the dispersed phase, which can be either water or oil. A quasi-static description successfully predicts a constant droplet size over a wide range of flow rates below some critical value.  The main distinction between our technique and the majority of prior techniques is that our outer fluid (the continuous phase) is not flowing, so the constriction and subsequent pinch-off of the dispersed phase is due to surface tension forces imposed by the shape of the class capillary rather than viscous forces.

\begin{acknowledgement}
Financial Support for this work was provided by National Sciences
and Engineering Research Council (NSERC).  The work of E.R.W. was
supported by the National Science Foundation under Grant No.
CBET-1336401.
\end{acknowledgement}


\end{document}